\documentclass[conference]{IEEEtran}

\usepackage{spconf,amsmath,graphicx}
\IEEEoverridecommandlockouts
\makeatletter
\def\thanks#1{\protected@xdef\@thanks{\@thanks
        \protect\footnotetext{#1}}}
\makeatother
\makeatletter
\renewcommand\footnoterule{\relax\kern-5pt
\hrule
\kern4.6pt}
\makeatother
\usepackage[nospace,noadjust]{cite}
\usepackage{amsmath,amssymb,amsfonts}
\usepackage{algorithmic}
\usepackage{graphicx}
\usepackage{textcomp}
\usepackage{xcolor}
\usepackage{subcaption}
\def\BibTeX{{\rm B\kern-.05em{\sc i\kern-.025em b}\kern-.08em
    T\kern-.1667em\lower.7ex\hbox{E}\kern-.125emX}}

\usepackage{algorithm,algorithmic}

\def\b1{{\boldsymbol{1}}}
\def\c1{{\textcircled{a}}}

\def\bb{{\boldsymbol{b}}}

\def\bn{{\boldsymbol{n}}}

\def\bx{{\boldsymbol{x}}}

\def\bz{{\boldsymbol{z}}}
\def\bA{{\mathbf{A}}}

\def\bC{{\boldsymbol{C}}}

\def\bF{{\mathbf{F}}}

\def\bI{{\mathbf{I}}}

\def\bS{{\boldsymbol{S}}}

\def\bZ{{\boldsymbol{Z}}}

\newtheorem{theorem}{Theorem}[section]

\begin{document}
%Copyright notice
%\pagestyle{empty}
\twocolumn[
\begin{@twocolumnfalse}

\large{\copyright  2024 IEEE.  Personal use of this material is permitted.  Permission from IEEE must be obtained for all other uses, in any current or future media, including reprinting/republishing this material for advertising or promotional purposes, creating new collective works, for resale or redistribution to servers or lists, or reuse of any copyrighted component of this work in other works.}\\
\\ %fill this for post-print

\end{@twocolumnfalse}
]
\title{Memory-Efficient Deep end-to-end Posterior Network (DEEPEN) for inverse problems}

%\title{Image recovery using Deep end-to-end Posterior Networks (DEEPEN)}

%\thanks{This work is supported by NIH R01-AG067078, R01-EB019961, and R01-EB031169.}
%}
\name{Jyothi Rikhab Chand,  Mathews Jacob}
\address{Department of Electrical and Computer Engineering, University of Iowa, IA, USA}

\maketitle

\begin{abstract}
End-to-End (E2E) unrolled optimization frameworks show promise for Magnetic Resonance (MR) image recovery, but suffer from high memory usage during training. In addition, these deterministic approaches do not offer opportunities for sampling from the posterior distribution. In this paper, we introduce a memory-efficient approach for E2E learning of the posterior distribution. We represent this distribution as the combination of a data-consistency-induced likelihood term and an energy model for the prior, parameterized by a Convolutional Neural Network (CNN). The CNN weights are learned from training data in an E2E fashion using maximum likelihood optimization. The learned model enables the recovery of images from undersampled measurements using the Maximum A Posteriori (MAP) optimization. In addition, the posterior model can be sampled to derive uncertainty maps about the reconstruction. Experiments on parallel MR image reconstruction
show that our approach performs comparable to the memory-intensive E2E unrolled algorithm, performs better than its memory-efficient counterpart, and can provide uncertainty maps. Our framework paves the way towards MR image reconstruction in 3D and higher dimensions. 
\end{abstract}


\section{Introduction}
Compressed Sensing (CS) algorithms have been widely used for the recovery of Magnetic Resonance (MR) images from highly undersampled measurements. These methods rely on an energy minimization formulation, where the cost function is the combination of data-consistency and regularization terms. The algorithms typically alternate between a data-consistency update and a proximal mapping step that can be viewed as a denoiser. Inspired by this, Plug-and-Play (PnP) methods \cite{pnp} replace the proximal operator with a Convolutional Neural Network (CNN), which is pre-trained as a denoiser for Gaussian noise corrupted images. End-to-End (E2E) training methods \cite{data} instead rely on algorithm unrolling to optimize the CNN for a specific forward model, thereby offering improvement in performance. However, the unrolling strategy requires a lot of memory during backpropagation, limiting the number of iterations. Deep Equilibrium Models (DEQ) \cite{deq, mol} were introduced to reduce memory demand by iterating a single layer until convergence to the fixed point. Both the DEQ and PnP methods require the CNN to be a contraction for theoretical convergence guarantees, which affects their practical performance.

All of the above deep learning methods can be viewed as Maximum A Posteriori (MAP) methods, differing mainly on how the prior is learned. On the contrary, we propose a memory-efficient Deep E2E learning Network (DEEPEN) to learn the posterior distribution, tailored for a specific forward model. This learned posterior model can be used to derive the MAP estimate, and samples can be generated to compute the uncertainty maps. Unlike fixed point PnP and DEQ methods that lack an energy-based formulation, we propose to represent the posterior distribution as the combination of a likelihood term determined by data consistency and a prior. We use an Energy-Based Model (EBM) for the prior \cite{energy_model}, which is parameterized by a CNN. Traditional EBM approaches \cite{energy_model} pre-learn the prior from training data, followed by their combination with a likelihood term for guidance. On the contrary, we directly learn the posterior for a specific forward model in an E2E fashion. We learn the CNN parameters using maximum likelihood optimization, which can be viewed as a contrastive learning strategy. In particular, training amounts to minimizing the energy for the \emph{true} reference samples, while the energy is maximized for the \emph{fake} samples drawn from the posterior. We use Langevin dynamics to generate the fake samples. Note that our approach does not involve algorithm unrolling, and therefore its memory demand during training is minimal. Unlike DEQ training, this approach does not require fixed-point iterations for backpropagation, which also reduces the computational complexity of the algorithm. Unlike existing models such as PnP and DEQ, our MAP estimation algorithm does not require Lipschitz constraints on the CNN for convergence, potentially enhancing performance. In contrast to DEQ and unrolled methods, the proposed posterior energy model can also be sampled to generate representative images and calculate uncertainty estimates. This approach is thus an alternative to diffusion models that use pre-trained CNN denoisers at various noise levels, which often require numerous iterations and specialized approaches to guide diffusion at different noise scales \cite{dps}. In comparison, our method uses a single-scale CNN, which leads to faster convergence.
\section{Posterior learning}
\begin{figure}[h]
\centering
    \includegraphics[trim={1cm 10cm 0cm 3cm},clip,width=0.5\textwidth]{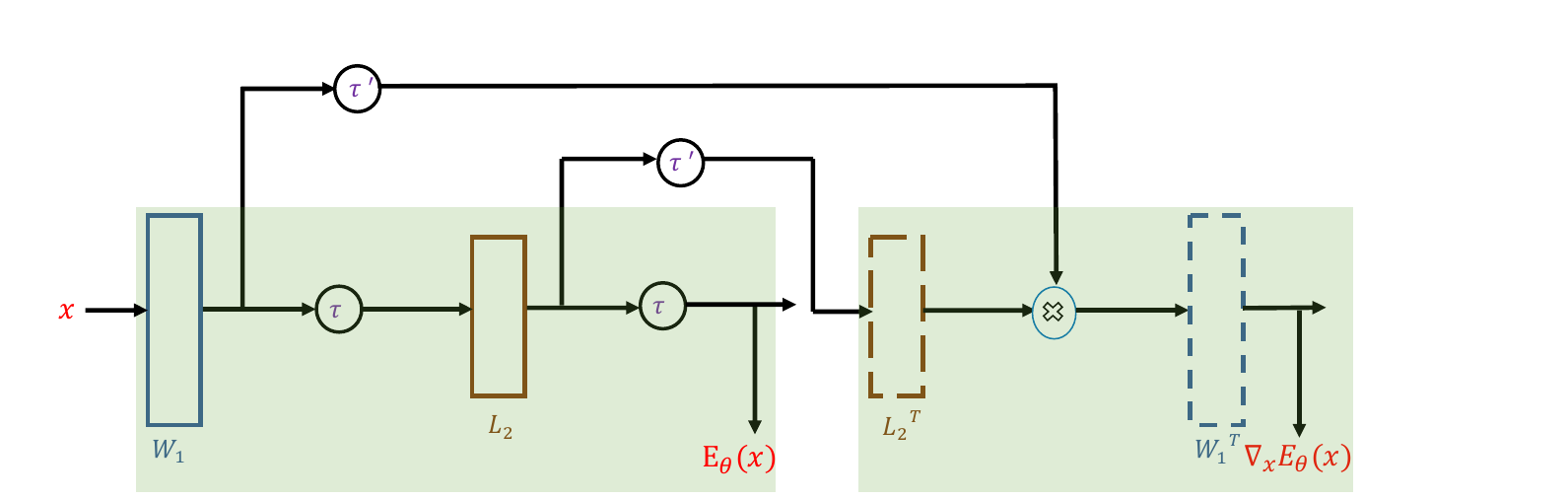}
    \caption{Illustration of computation of $E_{\boldsymbol \theta}(\bx)$ using a two-layer network. Its gradient $\nabla_\bx E_{\boldsymbol \theta}(\bx)$ is computed using the chain rule. $W_{1}$ and $W_{1}^T$ represents a convolutional and a transposed convolutional layer of appropriate size with shared weights, respectively; $L_{2}$ and $L_{2}^T$ are linear layers with shared weights; $\tau$ is the activation function and $\tau^{'}$ represents its gradient.  }
   
    \label{energy_network}
\end{figure}

We consider the recovery of an MR image $\bx \in \mathbb{C}^m$ from its corrupted undersampled measurements $\bb \in \mathbb{C}^n$:
\begin{equation}\label{linear}
   \bb = \bA\bx +\bn
\end{equation}
where $\bA \in \mathbb{C}^{n \times m}$ is known and $\bn \in \mathcal{N}(0,\eta^2\bI)$. 
The likelihood in this case is given by:
\begin{equation}
    p(\bb|\bx)=
\dfrac{1}{P}\exp\left(-\dfrac{\|\bA\bx-\bb\|^{2}}{2\eta^2}\right)
\end{equation}
where $P$ %=\int\exp\left(-\dfrac{1}{2}\|\bA\bx-\bb\|^{2}\right)d\bx$ 
is a normalizing constant. The negative log posterior of the distribution is specified by:
\begin{eqnarray}\label{eq:1}
-\log p(\bx|\bb) &=&\dfrac{1}{2\eta^2} \|\bA\bx-\bb\|_{2}^2  -\log p(\bx) + P
\end{eqnarray}
where $p(\bx)$ is the prior. 
%Then, the maximum a posteriori estimate of image $\bx$ is obtained by minimizing the negative log-posterior distribution:
%\begin{eqnarray}\label{eq:1}
%\bx^* &=& \arg \min_\bx\:-\log p(\bx|\bb)\\
      %&=&\arg \min_\bx \underbrace{\dfrac{1}{2\eta^2} \|\bA\bx-\bb\|_{2}^2}_{-\log p(\bb|\bx)}  + \underbrace{\phi(\bx)}_{-\lambda\log p(\bx)}\nonumber 
%\end{eqnarray}
We model the prior in \eqref{eq:1} as in \cite{energy_model}:
\begin{equation}\label{eq:2}
    p_{\boldsymbol \theta}(\bx) = \dfrac{1}{Z_{\boldsymbol \theta}}\exp(-E_{\boldsymbol \theta}(\bx))
\end{equation}
where $E_{\boldsymbol \theta}(\bx): \mathbb{C}^m \rightarrow \mathbb{R}^+$ is a neural network with $\boldsymbol \theta$ denoting its parameters. Any CNN model that takes an image and outputs a positive scalar value is sufficient for the above approach. We illustrate the energy model in Fig. \ref{energy_network} using a two-layer network. The gradient $\nabla_{\bx}E_{\boldsymbol \theta}(\bx): \mathbb{C}^m \rightarrow \mathbb{C}^m $ of the network can be evaluated using the chain rule. In the general case, the gradient operator can be derived using the built-in \emph{autograd} function. Note from Fig. \ref{energy_network} that the gradient resembles an auto-encoder with a one-dimensional latent space.  \\

\subsection{Maximum Likelihood training of the posterior }
\begin{figure}[t!]
\centering
    \includegraphics[trim={4cm 9cm 5cm 6cm},clip,width=0.5\textwidth]{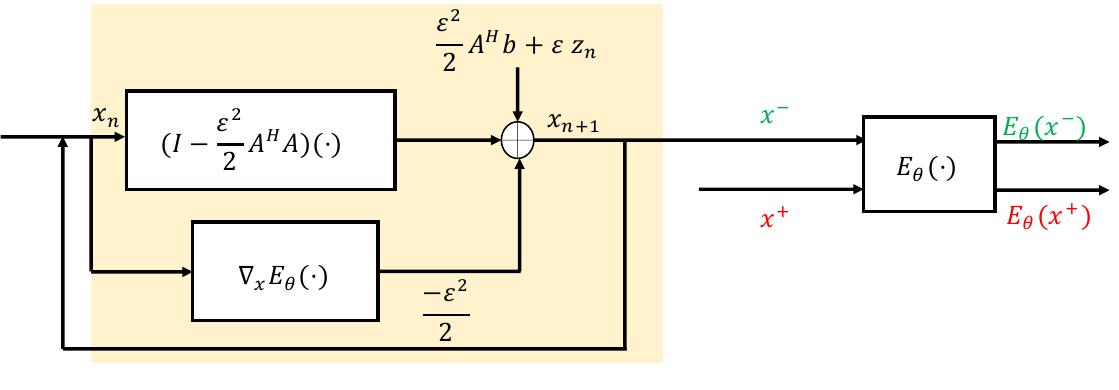}
    \caption{Illustration of training procedure of the proposed algorithm. The fake samples $\bx^-$ are generated using Langevin sampling, indicated by the yellow box. The intermediate results are not stored for backpropagation; so, a physical layer is sufficient for forward propagation, thus keeping the memory demand low. The samples $\bx^+$ are obtained from the training data. The training loss involves the energy difference between true and fake samples. Therefore, the training algorithm aims to modify the energy $E_{\boldsymbol \theta}(\cdot)$ so that the generated samples $\bx^-$ match the true samples $\bx^+$.}
    
\label{posterior_learning}
\end{figure}
Current EBM methods pre-learn $p(\bx)$ from the training data and then use it in \eqref{eq:1}, similar to PnP methods. Motivated by the improved performance of E2E methods, we learn the posterior distribution $p_{\boldsymbol\theta}(\bx|\bb)$ using the training data and for a specific forward model $\mathbf A$ in an E2E fashion using maximum likelihood optimization. In particular, we determine the optimal weights of the energy $E_{\boldsymbol \theta}(\bx)$ by  minimizing the negative log-likelihood of the training data set:
\begin{eqnarray}\label{eq:KL}
\boldsymbol \theta^* &=& \arg \min_{\boldsymbol \theta}\:\mathbb{E}_{\bx \sim q(\bx)}\Big[\underbrace{-\log p_{\boldsymbol \theta}(\bx|\bb)}_{\mathcal L_{\boldsymbol\theta}(\bx)}\Big]
\end{eqnarray}
where $q(\bx)$ is the probability distribution of the data and the posterior $p_{\boldsymbol \theta}(\bx|\bb)$ is defined as in (\ref{eq:1}) whose prior is defined in (\ref{eq:2}). We thus have:
\begin{eqnarray}
\label{eq:posterior_modeled}
\mathcal L_{\boldsymbol\theta}(\bx) &=&\dfrac{1}{2}\|\bA\bx-\bb\|^{2}+E_{\boldsymbol \theta}(\bx) +\log \tilde{\bZ}_{\boldsymbol \theta},
\end{eqnarray}
where $\tilde{\bZ}_{\boldsymbol \theta}=\int\exp\left(-\dfrac{1}{2}\|\bA\bx-\bb\|^{2}-E_{\boldsymbol \theta}(\bx)\right)d\bx$ is a normalizing constant. For simplicity, we absorb $\eta^2$ parameter into the definition of the energy.
% and the KL divergence is defined as \textcolor{red}{[ref]}:
% \begin{eqnarray}
% \rm{KL}(p(\bx|\bb) ||p_{\boldsymbol \theta}(\bx|\bb) )=\int p(\bx|\bb)\log p(\bx|\bb)d\bx \\\
% -\int p(\bx|\bb)\log p_{\boldsymbol \theta}(\bx|\bb)d\bx \\
% =-\int p(\bx|\bb)\log p_{\boldsymbol \theta}(\bx|\bb)d\bx+\rm{constant}
% \end{eqnarray}
% where the second equality is obtained by removing the term independent of $\boldsymbol \theta$. Expanding the term $-\int p(\bx|\bb)\log p_{\boldsymbol \theta}(\bx|\bb)d\bx$:
% \begin{equation}
%     \rm{KL}(p(\bx|\bb) ||p_{\boldsymbol \theta}(\bx|\bb) )=-\int p(\bx|\bb)\log p_{\boldsymbol \theta}(\bx|\bb)d\bx 
% \end{equation}
% \begin{equation}\label{eq:3}
% \begin{array}{ll}
% -\int p(\bx|\bb)\log p_{\boldsymbol \theta}(\bx|\bb)d\bx=\\\
% -\int p(\bx|\bb)\log \left(\exp
% \left(\dfrac{-\dfrac{1}{2}\|\bA\bx-\bb\|^{2}-E_{\boldsymbol \theta}(\bx)}{\tilde{\bZ}_{\boldsymbol \theta}}\right)\right)d\bx
% \end{array}
% \end{equation}
% where  The equation in (\ref{eq:3}) can be simplified as:
% \begin{equation}
% \begin{array}{ll}
% -\int p(\bx|\bb)\log p_{\boldsymbol \theta}(\bx|\bb)d\bx=\int p(\bx|\bb) \dfrac{\|\bA\bx-\bb\|^2}{2} d\bx + \\\
% \int p(\bx|\bb) E_{\boldsymbol \theta}(\bx) d\bx +\log \tilde{\bZ}_{\boldsymbol \theta}\\\
% =\mathbb{E}_{\bx \sim p(\bx|\bb)}\dfrac{\|\bA\bx-\bb\|^{2}}{2} +\mathbb{E}_{\bx \sim p(\bx|\bb)} E_{\boldsymbol \theta}(\bx)+\log \tilde{\bZ}_{\boldsymbol \theta}
% \end{array}
% \end{equation}
Consequently we have:
\begin{eqnarray}
  \nabla_{\boldsymbol \theta}\mathcal{L}_{\boldsymbol \theta}(\boldsymbol x)= \mathbb{E}_{\bx \sim q(\bx)} [\nabla_{\boldsymbol \theta} E_{\boldsymbol \theta}(\bx)] + \nabla_{\boldsymbol \theta}  \log \tilde{\bZ}_{\boldsymbol \theta}
\end{eqnarray}
The second term can be computed using the chain rule \cite{energy_model}:
\begin{eqnarray}
\nabla_{\boldsymbol \theta}  \log \tilde{\bZ}_{\boldsymbol \theta}=- \mathbb{E}_{\bx \sim p_{\boldsymbol \theta}(\bx|\bb)}[\nabla_{\boldsymbol \theta} E_{\boldsymbol \theta}(\bx)]
\end{eqnarray}
where $\bx \sim p_{\boldsymbol \theta}(\bx|\bb)$ are samples from the learned posterior distribution $p_{\boldsymbol \theta}(\bx|\bb)$. In the EBM literature, these are the generated or fake samples denoted by $\bx^-$, while the training samples are referred to as true samples, denoted by $\bx^+ \sim q(\bx)$. Thus, the ML estimation of $\boldsymbol{\theta}$ amounts to minimization of the loss:
\begin{eqnarray}\nonumber\label{final}
 \mathcal{L}'(\boldsymbol \theta)&=& \mathbb{E}_{\bx \sim q(\bx)} E_{\boldsymbol \theta}(\bx) -\mathbb{E}_{\bx \sim p_{\boldsymbol \theta}(\bx|\bb)} E_{\boldsymbol \theta}(\bx)\\\
  &\approx&  \left(\dfrac{1}{n} \displaystyle \sum_{i=1}^{n}{E_{\boldsymbol \theta}(\bx^{+}_i)}-\dfrac{1}{m}\displaystyle \sum_{j=1}^{m}{E_{\boldsymbol \theta}(\bx^{-}_j)}\right)
\end{eqnarray}
Intuitively, the training strategy (\ref{final}) will seek to decrease the energy of the true samples ($E_{\boldsymbol \theta}(\bx^+)$) and increase the energy of the fake samples ($E_{\boldsymbol \theta}(\bx^-)$). This may be seen as an adversarial training scheme similar to {\cite{gan}, where the classifier involves the energy itself as shown in Fig. \ref{posterior_learning}. The algorithm converges when the the fake samples are identical in distribution to the training samples; i.e, $E_{\boldsymbol \theta}(\bx^+) \approx E_{\boldsymbol \theta}(\bx^-)$.
\subsection{Generation of fake samples using Markov Chain Monte Carlo (MCMC)}\label{mcmc}
We generate the fake samples $\bx^- \sim p_{\boldsymbol \theta}(\bx|\bb)$ using Langevin MCMC method, which only requires the gradient of $\log p_{\boldsymbol \theta}(\bx|\bb)$ w.r.t. $\bx$ :
\begin{equation}\label{Ls}
\begin{array}{ll}
\bx_{n+1}=\bx_n  + \dfrac{\epsilon^2}{2}  \nabla_{\bx }\log p_{\boldsymbol \theta}(\bx_n|\bb)+\epsilon\bz_n \\\
\hspace{8mm}=\bx_n - \dfrac{\epsilon^2}{2} \left(\bA^H(\bA\bx_n-\bb)+\nabla_x E_{\boldsymbol \theta}(\bx_n)\right)+ \epsilon\bz_n 
\end{array}
\end{equation}
where $\epsilon>0$ is the step-size, $\bz_n \sim \mathcal{N}(0,\bI)$ and $\bx_0$ is drawn from a simple posterior distribution. The entire training process is summarized in Fig. \ref{posterior_learning}.
\subsection{Maximum aposteriori image recovery}
Once the posterior is learned, we obtain the MAP estimate by minimizing  (\ref{eq:posterior_modeled}) w.r.t. $\bx$ using the classical gradient descent algorithm:
\begin{equation}\label{gd_up}
\bx_{k+1} = \bx_k - \alpha_k \left(\bA^H(\bA\bx_k-\bb)+\nabla_x E_{\boldsymbol \theta}(\bx_k)\right)
\end{equation}
where $\alpha_k$ is the step-size and is found using the backtracking line search method \cite{Wright}.  This line search method starts with a large step-size value, i.e., $\alpha_{k}=1$, and keeps decreasing it by a factor of $\beta \in (0,1)$ until the chosen step-size sufficiently decreases the cost function. The decrease in the cost function is typically measured using the Armijo–Goldstein rule, which is $\mathcal L_{\theta}\left(\bx_k - \alpha_k \nabla_x \mathcal L_{\theta}(\bx_k)\right) > \mathcal L_{\theta}(\bx_k) -\beta \alpha_k \|\nabla_x \mathcal L_{\theta}(\bx_k)\|_{2}^{2}$, thereby ensuring that the chosen step-size monotonically decreases the cost function at every iteration. The following result shows that the proposed algorithm converges monotonically to a local minimum of the cost function (\ref{eq:posterior_modeled}). 
\begin{theorem}[\cite{Wright}]\label{l1}
Consider the cost function $\mathcal L_{\theta}(\bx)$ in (\ref{eq:posterior_modeled}), which is bounded below by zero\footnote{The CNN implementation $E_\theta(\bx)$ has a Rectified Linear Unit (RELU) in the output layer, which makes the lower bound zero.}. Then, the steepest descent optimization scheme in \eqref{gd_up} with backtracking line search to determine the optimal step-size $\alpha$ will monotonically decrease the cost function (i.e., $\mathcal L_{\theta}(\bx_k+1) \leq \mathcal L_{\theta}(\bx_k)$). In addition, the sequence $\{\bx_k\}$ will converge to a stationary point of (\ref{eq:posterior_modeled}), provided the gradient $\nabla_\bx \mathcal L_{\theta}(\bx)$ is Lipschitz-continuous.  
\end{theorem}
% The pseudocode of the algorithm is shown below. 
% \begin{center}
% \begin{tabular}{@{}p{9cm}}
% \hline
% \hline
% \bf{Algorithm1: Gradient descent with backtracking line search} \\
% \hline
% \hline
% {\bf{Inputs}}: Forward operator $\bA$, undersampled measurements $\bb$, pre-trained $E_{\boldsymbol \theta}(\bx)$,$\gamma=\beta=0.5$ \\
% {\bf{Initialize}}: Initialize $\bx_{0}$.  \\
% {\bf{Repeat}}: Given $\bx_k$ perform the $k+1$-th step. \\
% \hspace{3mm}Set $\alpha_k=1.$\\
% \hspace{3mm}Compute $\bz=\bx_k - \alpha_k \left(\bA^H(\bA\bx_k-\bb)+\nabla_x E_{\boldsymbol \theta}(\bx_k)\right)$.\\
% \hspace{3mm} \textbf{while} $\mathcal L_{\theta}(\bz)>\mathcal L_{\theta}(\bx_k)-\beta\alpha_k\|\nabla_x \mathcal L_{\theta}(\bx_k)\|_{2}^{2}$:\\
% \hspace{8mm}$\alpha_k=\gamma\alpha_k$\\
% \hspace{8mm}Re-compute $\bz=\bx_k - \alpha_k \left(\bA^H(\bA\bx_k-\bb)+\nabla_x E_{\boldsymbol \theta}(\bx_k)\right)$.\\
% \hspace{3mm}$\bx_{k+1 }= \bz$\\
% \hspace{3mm}$k\leftarrow k+1$, {\bf{until convergence}}\\
% {\bf{Output}}: $\bx_{\rm{MAP}}^{*} = {\bx}_{k}$
% \\
% \hline
% \hline
% \end{tabular}
% \end{center}

%The proof is omitted due to space constraints but can be found in \cite{Wright}. Assuming that the maximum eigen-value of $\bA^H\bA$ is one, then $\nabla_\bx f(\bx)$ is Lipschitz continuous with Lipschitz constant $1+L$, where $L$ is the Lipschitz constant of $E_\theta(\bx)$ and can be computed using \cite{CLIP, SN}. Therefore, by Theorem. \ref{l1}, the gradient descent will converge to a stationary point of the problem (\ref{map}).\\
Note that the above theorem is applicable irrespective of the Lipschitz constant of the CNN, and when $\nabla_\bx \mathcal L_{\theta}(\bx)$ is a conservative vector field i.e., it is the gradient of $\mathcal L_{\theta}(\bx)$.
\begin{figure}[t!]
        \centering
        \begin{subfigure}[b]{0.475\textwidth}
            \centering
            \includegraphics[width=\textwidth]{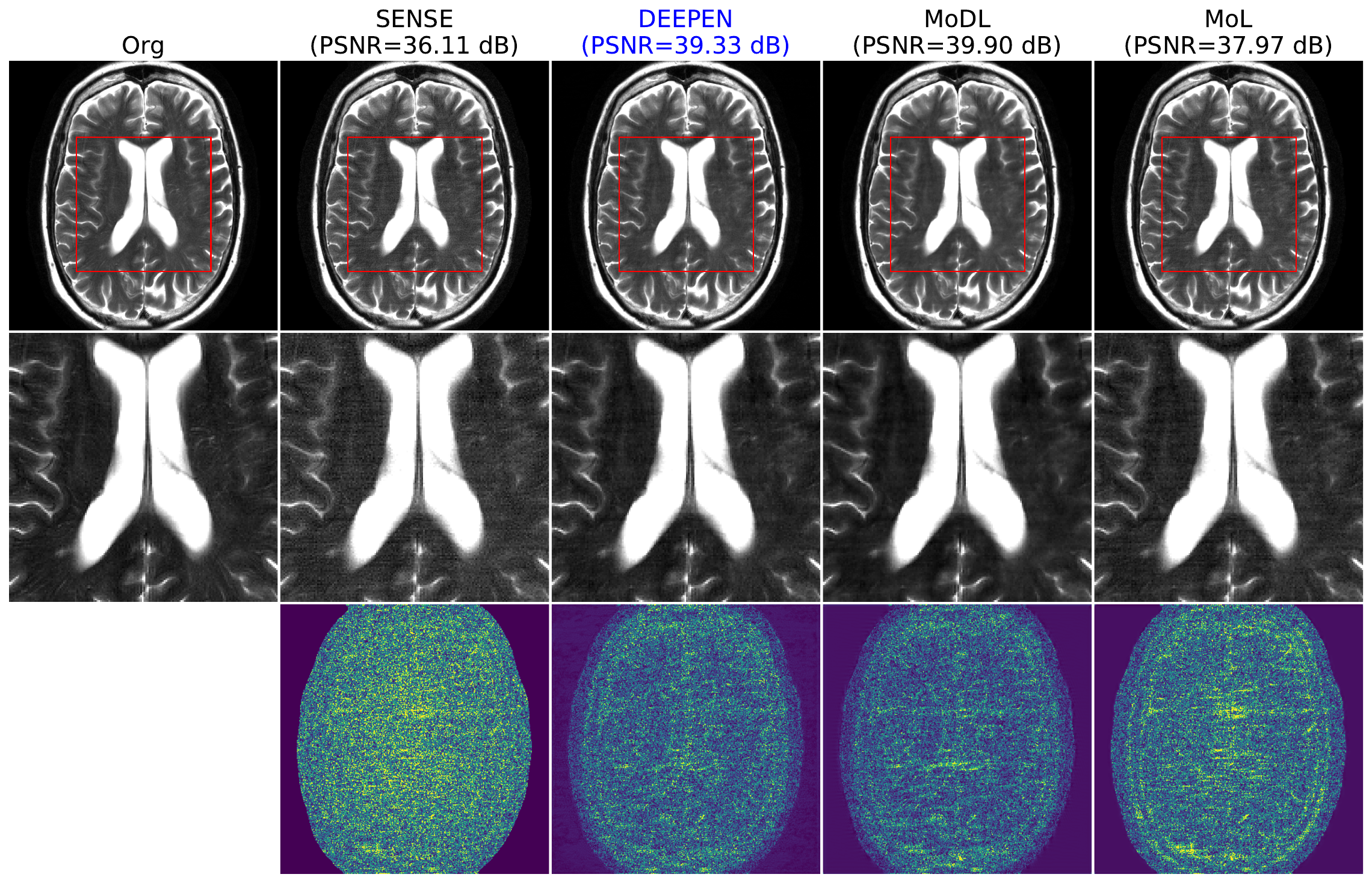}
            \caption{\small T2}  
            \label{T2}
        \end{subfigure}
        \hfill
        \begin{subfigure}[b]{0.475\textwidth}  
            \centering 
            \includegraphics[width=\textwidth]{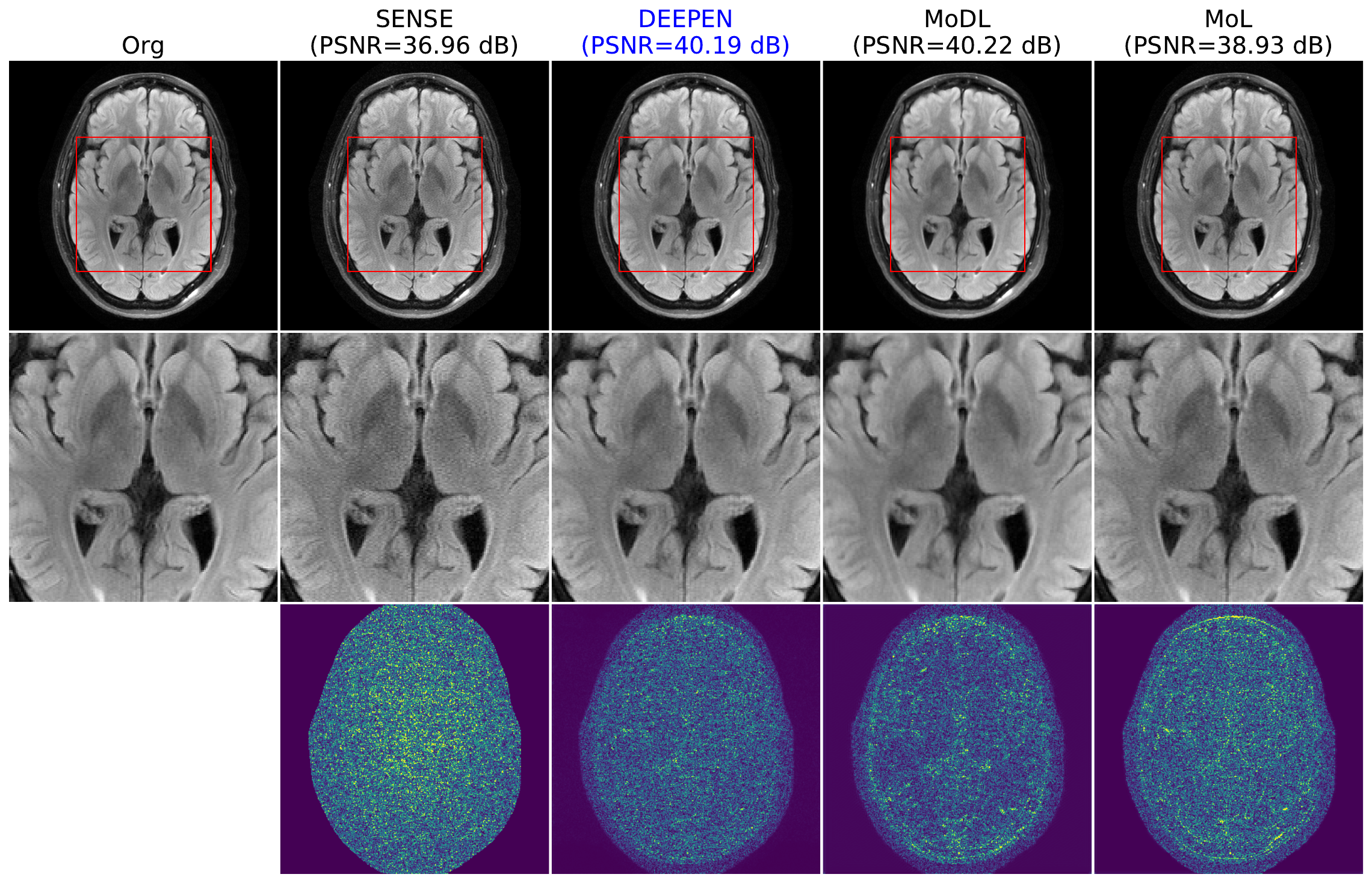}
            \caption{\small FLAIR}   
            \label{Flair}
        \end{subfigure}
        \caption{Comparision of DEEPEN with MoDL and MoL for two different contrasts: (a) T2 and (b) FLAIR. Top row shows the reconstructed image, second row shows the enlarged image, and the third row is the error image.} 
        \label{different_contrast}
    \end{figure}
\section{Experiments}\label{results}
\subsection{Data set}
We evaluated the performance of DEEPEN on the publicly available parallel fastmri brain data set which contains FLAIR and T2-weighted images \cite{brain_dataset}. It is a twelve-channel brain data set and consists of complex images of size $320 \times 320$. The matrix $\bA$ in (\ref{linear}) in this case is $\bA=\bS\bF\bC$, where $\bS$ is the sampling matrix, $\bF$ is the Fourier transform and $\bC$ is the coil sensitivity map which is estimated using ESPIRiT algorithm \cite{espirit}. The data set for each contrast was split into $45$ training, $5$ validation, and $50$ test subjects. 
DEEPEN was trained and evaluated on both contrasts separately for a four-fold retrospective undersampled measurement, which was obtained by sampling
the data along the phase-encoding direction using a 1D non-uniform variable density mask.
\subsection{Architecture and implementation }
The network $E_{\boldsymbol \theta}(\bx)$ defined in (\ref{eq:2}) was built using five 3x3 convolutional layers with 64 channels each, followed by a linear layer. A ReLU was used between each convolutional layer and at the end of the linear layer. We evaluated the gradient $\nabla_\bx E_{\boldsymbol \theta}(\bx)$ using the chain rule. 

The MCMC sampling was performed for $30$ iterations and was initialized with $\bx_0=(\bA^H\bA +\tilde{\lambda}\bI)^{-1} \bA^{H}\bb$ which is obtained by minimizing the negative logarithmic posterior distribution $-\log p_0(\bx|\bb)=\|\bA\bx-\bb\|_{2}^{2}+\tilde{\lambda}\|\bx\|_{2}^2$ w.r.t. $\bx$. Next, 
similar to \cite{anatomy}, for stable training we found it beneficial to: a) add Gaussian noise of standard deviation $2\epsilon$ to the training data $\{\bx_i^{+}\}$ that effectively made the training data smooth, and b) scale $\nabla_x\log p_{\boldsymbol \theta}(\bx|\bb)$ by $\dfrac{\epsilon^{2}}{2}$ which is equivalent to using the following Langevin MCMC update:
\begin{equation}\
\begin{array}{ll}
\bx_{n+1}=\bx_n -  \left(\bA^H(\bA\bx_n-\bb)+\nabla_x E_{\boldsymbol \theta}(\bx_n)\right)+ \epsilon\bz_n 
\end{array}
\end{equation}
DEEPEN was trained with $\epsilon =0.001$ and the optimal $\boldsymbol \theta$ was found using the Adam optimizer. The gradient descent algorithm was run until
$\dfrac{|\mathcal L_{\boldsymbol\theta}(\bx_{k+1})-\mathcal L_{\boldsymbol\theta}(\bx_k)|}{|\mathcal L_{\boldsymbol\theta}(\bx_{k})|} \leq 10^{-6}$.\\
We compared the performance of the proposed algorithm with SENSE \cite{sense}, MoDL \cite{data}, and MoL \cite{mol}. The unrolled algorithms were trained in an E2E fashion for $10$ iterations. A five-layer CNN was used for both the unrolled algorithms. The Lipschitz constraint in case of MoL was implemented using the log-barrier approach.  
% \begin{table}[H]
% \centering
% \caption{Comparison of Avg.PSNR of the end-to-end models for four different contrast types (\textcolor{red}{capital??})}
% \begin{tabular}{|p{1cm}|p{2cm}|p{2cm}|p{2cm}|}
% \hline
% Contrast& Proposed Method & MoDL&MoL \\
% \hline
% T1post &42.04   &42.99 &39.47 \\
% \hline
% FLAIR&37.2 & 37.97 &36.31 \\
% \hline
% T2&39.17  & 40.13  & 38.12 \\
% \hline
% T1 &37.61   & 38.73 &36.46\\
% \hline
% \end{tabular}
% \label{Comp}
% \vspace{-1em}
% \end{table}
\section{Results}
\begin{figure}[t!]
\centering
    \includegraphics[width=0.5\textwidth]{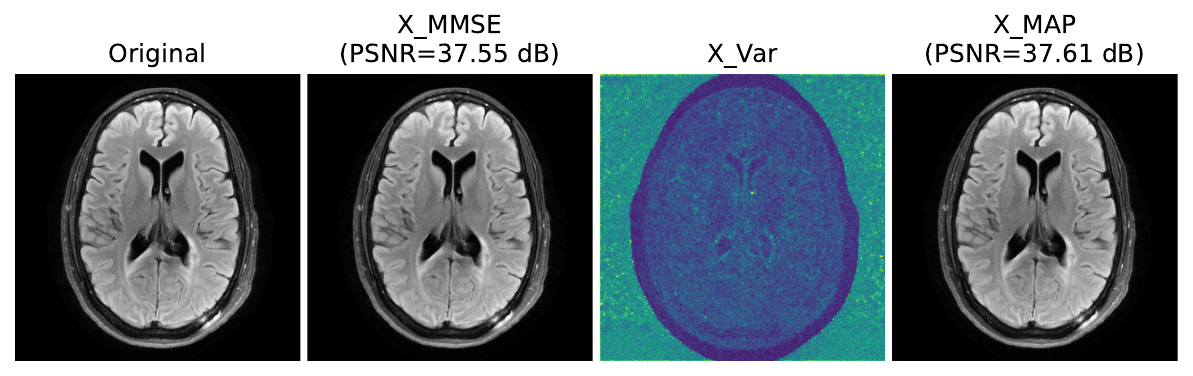}
    \caption{MMSE, uncertainty and the MAP estimate given by the DEEPEN algorithm on the four-fold undersampled FLAIR image. The MMSE and the uncertainty map was obtained by taking the mean and variance over $100$ samples. }
\label{mmse}
\end{figure}
\subsection{Maximum aposteriori estimates}
Table \ref{cmp_table} compares the performance of the reconstruction algorithms on the test data. Fig. \ref{T2} and Fig. \ref{Flair} show the reconstructed images of T2-weighted and FLAIR images, respectively. PSNR and  SSIM  are used as evaluation metrics. Table \ref{cmp_table} shows that DEEPEN and MoDL perform better than SENSE and MoL, and perform comparably with each other. While SENSE is a CS-based algorithm and does not have a neural network module, the lower performance of the unrolled MoL algorithm is due to the Lipschitz constraint on its CNN. This constraint is required to ensure convergence of MoL to the fixed point. Note that, according to Theorem 2.1, the DEEPEN algorithm can ensure convergence to a stationary point without any constraint on  $\nabla_{x}E_{\boldsymbol \theta}(\bx)$ - which results in improved performance. We would like to remind the reader that although DEEPEN and MoDL have similar performance, DEEPEN is memory-efficient and hence paves the way to reconstruct images of higher dimensions (for example, 3D brain volumes) which is not possible with MoDL. 

\begin{table}[H]
\centering
 \caption{Comparison of DEEPEN MAP estimates against SENSE, MoDL and MoL for two different 4-fold undersampled contrast types.}
\begin{tabular}{|p{1.3cm}|p{1.4cm}p{1cm}|p{1.4cm}p{1cm}| }
    \hline
{Contrast}& \multicolumn{2}{c|}{FLAIR}& \multicolumn{2}{c|}{T2}
                                       \\                 
\cline{1-5}
 &  Avg. PSNR(dB) &Avg. SSIM&Avg. PSNR(dB) &Avg. SSIM\\
\hline
SENSE&34.24&0.931&36.14&0.95\\
\hline
\textbf{DEEPEN} &  \textbf{37.14} & \textbf{0.96}  & \textbf{39.22} & \textbf{0.98}     \\
\hline
MoDL& 37.97  &   0.97 &40.13&0.98\\
\hline
MoL&36.31  &  0.96 &38.12& 0.97\\
\hline
\end{tabular}
\label{cmp_table}
\end{table}
\subsection{Bayes estimation}
Given the posterior distribution $p_{\boldsymbol\theta}(\bx|\bb)$, we can now draw samples from it. This aids in estimating the Minimum Mean Square Error (MMSE) and the uncertainty map of the reconstructed image. Fig. \ref{mmse} shows the MMSE and the uncertainty map provided by DEEPEN on a four-fold undersampled FLAIR-contrast MR brain image. The MMSE and the uncertainty map are estimated by computing the mean and variance of $100$ samples from the posterior distribution. The samples are obtained using the Langevin MCMC method initialized randomly from a Gaussian distribution. 
\section{Conclusion}
In this paper, we proposed a memory-efficient E2E training framework DEEPEN, which does not require constraining the Lispchitz constant of the CNN. Consequently, DEEPEN achieved a better PSNR than the memory-efficient unrolled algorithm MoL. Moreover, we have a well-defined cost function that allows guaranteed convergence to a stationary point and enables the use of faster sampling algorithms such as Metropolis-Hastings to obtain the uncertainty map of the estimate. This work can be extended to MR image reconstruction in higher dimensions.

% we represented the proposed posterior distribution as an EBM. The training procedure involved minimizing the energy difference between the true and the fake samples. The fake samples were generated by unrolling the update steps of  Langevin sampling but did not require storing the gradients, making the training process memory efficient. Also, the Langevin sampling involved the data-consistency term, and consequently, similar to unrolled algorithms, the training of the proposed approach is done in an end-to-end fashion. Moreover, unlike MoL (the memory-efficient unrolled algorithm), we did not constrain the Lipschitz of the CNN. Consequently, as illustrated in Table. I, the proposed approach DEEPEN, achieves a better PSNR than MoL. Theorem 1 also showed that the iterates of the proposed algorithm is guaranteed to converge to a stationary point of the negative log-posterior distribution. 

\section{Compliance with ethical standards}
\label{sec:ethics}

This study was conducted on a publicly available human subject data set. Ethical approval was not required, as confirmed by the license attached with the open-access data.

\section{Acknowledgments}
\label{sec:acknowledgments}

This work is supported by NIH grants R01-AG067078, R01-EB031169, and R01-EB019961.
\bibliographystyle{IEEEtran}
\bibliography{ref}
\end{document}